\documentclass[a4paper]{article}

\usepackage[pages=all, color=black, position={current page.south}, placement=bottom, scale=1, opacity=1, vshift=5mm]{background}

\usepackage[margin=1in]{geometry} 

\usepackage{amsmath}
\usepackage{amsthm}
\usepackage{amssymb}
\usepackage{bm}

\usepackage[utf8]{inputenc}
\usepackage{hyperref}
\hypersetup{
	unicode,
	pdfauthor={Antonio Tripodo},
	pdftitle={Mutual Information in Molecular and Macromolecular Systems},
	pdfsubject={Mutual Information in Molecular and Macromolecular Systems},
	pdfkeywords={glass transition; mutual information; Molecular Dynamics; dynamical heterogeneity; Johari--Goldstein relaxation; roto-translation coupling},
	pdfproducer={LaTeX},
	pdfcreator={pdflatex}
}

\usepackage[sort&compress,numbers,square]{natbib}
\theoremstyle{plain}

\theoremstyle{definition}

\usepackage{graphicx, color}
\graphicspath{{fig/}}

\usepackage{algorithm, algpseudocode} 
\usepackage{mathrsfs} 

\usepackage{lipsum}

\title{Mutual Information in Molecular and Macromolecular Systems}
\author{Antonio Tripodo$^1$\and Francesco Puosi$2$ \and Marco Malvaldi$^1$ \and Dino Leporini$^{1,3}$}

\date{
	$^1$Dipartimento di Fisica ``Enrico Fermi'', Universit\`a di Pisa, Largo B.\@Pontecorvo 3, I-56127 Pisa, Italy \\ \texttt{antonio.tripodo@df.unipi.it (A.T.); marcoampelio@hotmail.com (M.M.)}\\%
	$^2$ Istituto Nazionale di Fisica Nucleare (INFN), Sezione di Pisa, Largo B.\@Pontecorvo 3, I-56127 Pisa, Italy; \\ \texttt{ francesco.puosi@pi.infn.it}\\[2ex]%
	$^3$ Istituto per i Processi Chimico-Fisici-Consiglio Nazionale delle Ricerche (IPCF-CNR), Via G. Moruzzi 1, I-56124 Pisa, Italy \\ \texttt{ dino.leporini@unipi.it}\\[2ex]%
}

\begin{document}
	\maketitle
	
	\begin{abstract}
		The relaxation properties of viscous liquids close to their glass transition (GT) have been widely characterised by the statistical tool of time correlation functions. However, the strong influence of ubiquitous non-linearities calls for new, alternative tools of analysis. In this respect, information theory-based observables and, more specifically, mutual information (MI) are gaining increasing interest.
		Here, we report on novel, deeper insight provided by MI-based analysis of molecular dynamics simulations of molecular and macromolecular glass-formers on two distinct aspects of transport and relaxation close to GT, namely dynamical heterogeneity (DH) and secondary Johari--Goldstein (JG) relaxation processes. In a model molecular liquid with significant DH, MI reveals two populations of particles organised in clusters having either filamentous or compact globular structures that exhibit different mobility and relaxation properties.
		In a model polymer melt, MI provides clearer evidence of JG secondary relaxation and sharper insight into its DH. It is found that both DH and MI between the orientation and the displacement of the bonds reach (local) maxima at the time scales of the primary and JG secondary relaxation. This suggests that, in (macro)molecular systems, the mechanistic explanation of both DH and relaxation must involve rotation/translation coupling.
		\end{abstract}
		
		
		\section{Introduction}
		\label{Intro_Gen}
		The nature of the solidification process observed at the glass transition (GT) temperature {$T_{\mathrm{g}}$} by cooling supercooled viscous liquids is a topic of intense research. However, a clear characterisation of all the phenomena related to the glass transition as well as a deep microscopic understanding of the glassy state remains an open challenge \cite{DebenedettiBook,DebeStilli2001,BerthierBiroliRMP11}. 
		{
			Both frequency response and time relaxation of viscous liquids close to GT exhibit clear {\it {nonlinear} 
			}features, thus motivating intense experimental \cite{SchmidtRohrSpiessPRL91,BohmerScience96,BohmeEPL96}, simulation  \cite{GlotzerJCP03,ReichmannPRL07}, and theoretical \cite{BerthierBiroliRMP11} research. Within the framework of the familiar {\it linear} response, the fluctuation--dissipation theorem evidences that  two-time correlation functions incorporate all of the information to evaluate the susceptibility of a system in thermal equilibrium \cite{SethnaCorrFunctBook}. In principle, as outlined by Kubo in his seminal paper on adiabatic linear response theory, a formal treatment of the adiabatic nonlinear response in terms of a series expansion involving two-, three-, four-, and higher-order time correlation functions may be developed \cite{Kubo1957}. However, these expansions are exceedingly difficult to translate into a useful, experimentally verifiable form and are nowadays believed to be not a viable approach for most transport processes and irreversible relaxation processes \cite{evans_morriss_2008}. This poses the question of alternative approaches to revealing nonlinear aspects of relaxation and response.
			
			Starting with Galton’s first expositions in 1888 on how to accurately characterise the dependence (correlation) between two or more random variables, a plethora of measures have been developed \cite{SmithMI_NonLin15}. Perhaps, the most popular  is the Pearson correlation coefficient:
			\begin{equation}
				C(X,Y)=\frac{\langle(x-\langle x\rangle)(y-\langle y\rangle)\rangle}{\sigma_X\sigma_Y}
				\label{Pearson_def}
			\end{equation}
			where $\sigma_X$ is the standard deviation of $X$. A disappointing issue of the Pearson coefficient, however, is that it measures only {\it linear} dependence.

			To overcome this limitation, mutual information (MI) attracted quite a large interest. MI is defined as follows \cite{MezardMontanariInformationTheory}:
			\begin{equation}
				I(X,Y) = \int \int \mathrm{d}x\; \mathrm{d}y \; p(x,y)\log \left[\frac{p(x,y)}{p(x)p(y)}\right] 
				\label{mut_inf}
			\end{equation}
			where $p(x,y)$ is the joint probability distribution of the random variables $X$ and $Y$ with distributions $p(x)$ and $p(y)$, respectively.  Two random variables with no MI are independent. 
			MI can be thought of as the reduction in uncertainty (ignorance \cite{SethnaCorrFunctBook}) about one random variable given knowledge of another  \cite{LathamShortIntro_MI09,SmithMI_NonLin15}. Interestingly, the data processing inequality states that manipulation of the data does not increase MI \cite{CoverThomasInfoTh06}. 
			
			Even if developed in information theory and applied to maximise the amount of information transferred in communication, the notion of MI extended influence to many other fields \cite{CoverThomasInfoTh06}. A basic motivation is the property that MI assumes {\it no} sort of underlying distribution or dependence between random variables \cite{Li:1990sf}. In particular, it is known that the nonlinear dependence leads to rather subtle effects \cite{SmithMI_NonLin15}. In this respect, Figure \ref{circ_dist} offers further illustration.
			
			The sensitivity of MI to nonlinear dependence is certainly appealing in studies on GT \cite{Dunleavy12,Dunleavy:2015fq,Tong_Tanaka_PRL2020} as well as in other fields of physics, including topological transition in the XY model \cite{Iaconis:2013ij}, phase transition in a 2D disordered Ising model  \cite{Sriluckshmy:2018dn} and evaluation of the configurational entropy of liquid metals \cite{Gao:2018fv}. On the other hand, unlike the Pearson correlation coefficient, the MI evaluation requires knowledge of the distributions of random variables. If these are not known, their estimation is a delicate (and sometimes uncertain) procedure leading to higher computational cost \cite{MezardMontanariInformationTheory,CoverThomasInfoTh06}.

			\begin{figure}[h]
				\centering
				\includegraphics[width=0.7\textwidth]{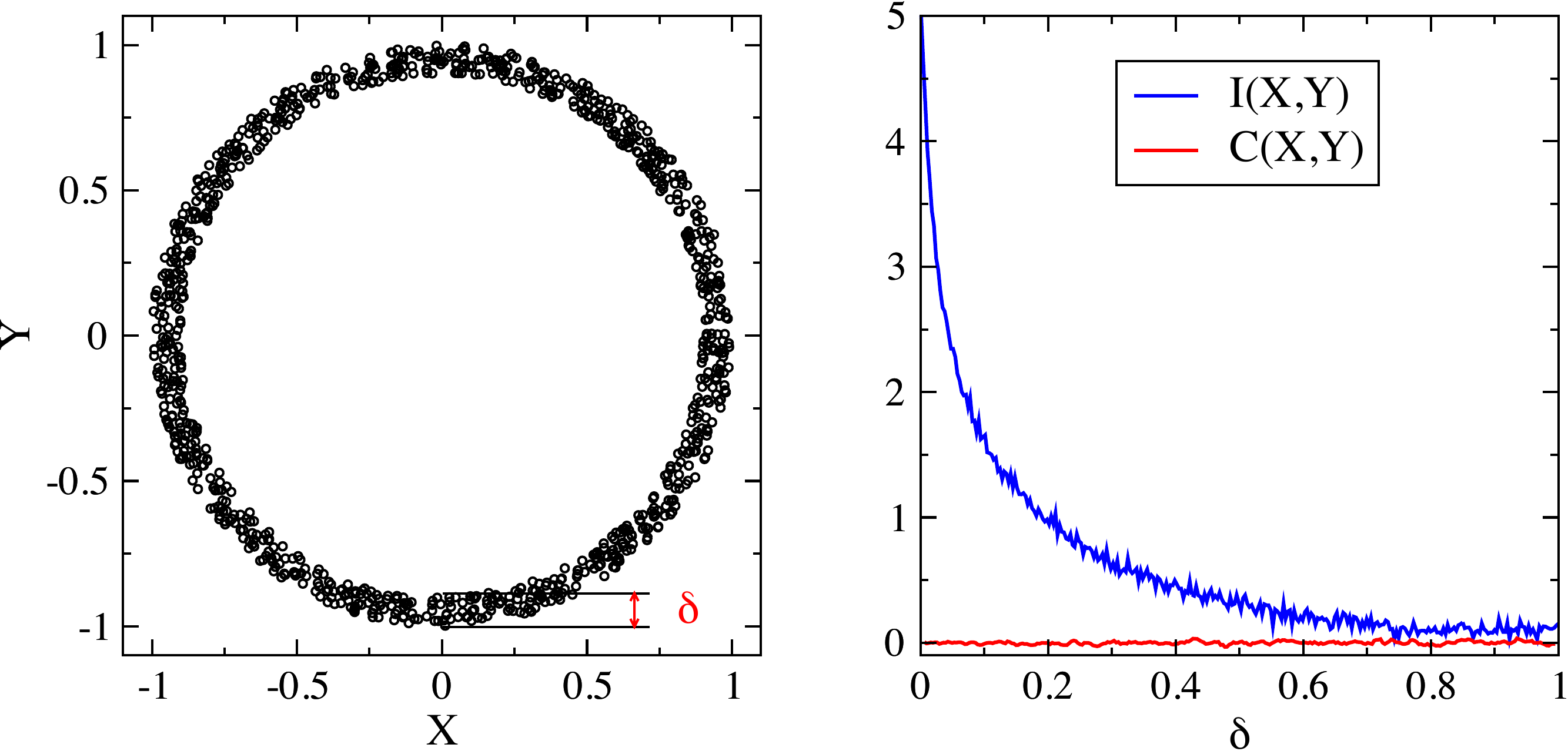}	
				\caption{{Illustration} 
					of the sensitivity of mutual information $I(X,Y)$ to detecting the nonlinear dependence between two random variables $X$ and $Y$ compared with the Pearson correlation coefficient $C(X,Y)$. {(\textbf{Left})} sample of dots $(X,Y)$ with uniform distribution over an annulus of mean radius $r_0=1$ and thickness $\delta$. {(\textbf{Right})} mutual information and Pearson correlation coefficient of samples with different thickness. No correlation is found by the Pearson  coefficient.}
				\label{circ_dist}
			\end{figure}
			
			Building on previous studies carried out by Molecular Dynamics (MD) simulations, the present paper reports novel insight provided by MI-based approaches on two distinct aspects of transport and relaxation close to GT, namely dynamical heterogeneity \cite{Tripodo_SM2019,Tripodo_epje_2019}  and its influence on secondary relaxations \cite{Tripodo_polymers2020,PuosiTripodo_Macromolecules2021}.  Details about the model systems and numerical methods are found elsewhere \cite{Tripodo_SM2019,Tripodo_epje_2019,Tripodo_polymers2020,PuosiTripodo_Macromolecules2021}.
			
		}
		\section{{Dynamical Heterogeneity and Mutual Information between Particle Displacements}}
		
		The transition from a liquid to a glass is accompanied by the growth of transient domains that exhibit different mobility. This phenomenon is usually dubbed ``dynamical heterogeneity'' (DH) and has extensively been studied \cite{SILLESCURevDynHet99, Ediger00, Richert02, BerthierBiroliRMP11}.
		The size of the DH domains is relatively small, involving about 10 molecule diameters \cite{Ediger00},  corresponding to a few nanometers \cite{TrachtSpiessDynHetPRL98} and strictly related to the possible presence of characteristic length scales in glass-forming systems \cite{AdamGibbs65}. 
		Even if growing static length scales have been reported by experiments \cite{BiroliBouchaudLoidlLunkenheimerScience16} and simulations \cite{BiroliKarmakarProcacciaPRL13}, there is still debate about whether they control the glass transition \cite{CatesWyartPRL17}. 
		On the other hand, it is not clear to what extent dynamic correlations are a consequence or the primary origin of the slow dynamics occurring close to the GT \cite{SastryLengthScalesRepProgrPhys15}.

		In a pioneering work in 2015, Dunleavy et al.  \cite{Dunleavy:2015fq} proposed a MI approach to investigate DH in a polydisperse mixture of hard spheres. They considered the MI between pairs of particles displacements, obtained a dynamic length scale growing on approaching GT, and revealed two distinct fractions of particles, so-called \textit{early} and \textit{late} fractions. 
		{
			To test its robustness, we extended the MI approach of Reference \cite{Dunleavy:2015fq}  to the context of {\it molecular liquids} and investigated by MD simulations \cite{Tripodo_SM2019,Tripodo_epje_2019}. For readers' convenience, these previous results are briefly outlined in Section \ref{previousDH}. Then, Section \ref{clustering} presents novel results proving that the so-called \textit{early} and \textit{late} populations are arranged in localised clusters. 
		}
		
		\subsection{Mutual Information Reveals DH in a Molecular Liquid}
		\label{previousDH}

\begin{figure}[h]
			\centering
			\includegraphics[width=0.8\textwidth]{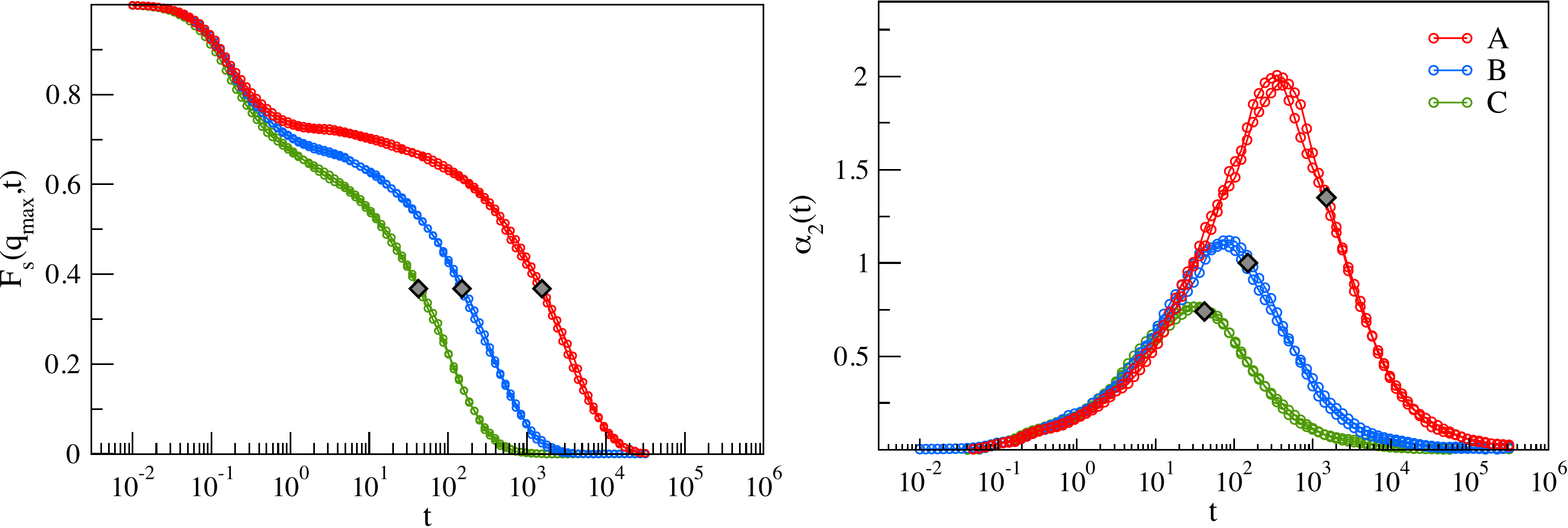}
			\caption{{(\textbf{Left}) {ISF} from Equation (\ref{isf}) at $q_{max}$, the first peak of the static structure factor. The curves refer to six states designed to have ISFs grouped into three distinct pairs, labelled as A, B, and C; see Reference {\cite{Tripodo_SM2019}} 
					for details. In terms of structural relaxation time, density, and temperature $(\tau_\alpha,\rho, T)$, each set is identified as A = $\{(42, 1.01, 0.47), (42, 1.05, 0.6) \}$, B = $\{(150, 1.01, 0.435), (150, 1.03, 0.49) \}$, \linebreak C = $\{(1500, 1.02, 0.42), (1500, 1.05, 0.51) \}$. (\textbf{Right}) NGP from Equation (\ref{NGP}), of the same six states showing that their NGP group likewise. The relaxation time $\tau_\alpha$ is marked by a diamond in both plots. The data were taken from Reference \cite{Tripodo_SM2019}.}}
			\label{ISF+NGP}
		\end{figure}

		MD simulations of a model molecular liquid made of fully flexible trimers have been carried out with technical details given elsewhere  \cite{Tripodo_SM2019,Tripodo_epje_2019}. Relaxation and heterogeneous transport are characterised by the intermediate scattering function (ISF) and  the non-Gaussian parameter (NGP) \cite{HansenMcDonaldIIIEd}, respectively. ISF  is defined as
		\begin{equation}
			F_s({\bf q},t)=\frac{1}{N} \sum_{j=1}^{N}\left \langle e^{ i {\bf q}\cdot[{\bf r}_j(t)-{\bf r}_j(0)]} \right \rangle
			\label{isf}
		\end{equation} 
		where ${\bf r}_j(t)$ is the position of the jth particle at time $t$. The brackets denote suitable ensemble averages, and $N$ is the total number of particles.
		In an isotropic liquid, ISF depends only on the modulus of the wavevector $q = || \bm q ||$ and provides a convenient relaxation function to study the rearrangements of the spatial structure of the fluid over the length scale $\sim 2\pi/q$. We define the structural relaxation time $\tau_{\alpha}$ by the relation $F_s(q_{max}, \tau_{\alpha}) = e^{-1}$ where $q_{max}$ is the maximum of the static structure factor. In the present system, $q_{max} \simeq 2 \pi/\sigma$, with $\sigma$ being roughly the particle diameter. NGP is defined as follows:
		\begin{equation}
			\alpha_2(t) = \frac{3\langle\delta r^4(t)\rangle}{5\langle\delta r^2(t)\rangle^2} - 1
			\label{NGP}
		\end{equation}
		$\alpha_2(t)$ vanishes if the particle displacement $\delta \vec{r}(t)$ is Gaussian. NGP is a well-known DH metrics \cite{DebenedettiBook,BerthierBiroliRMP11,Richert02}.
		By suitably adjusting both the density $\rho$ and the temperature $T$ of the liquid, six states were prepared to be grouped in three pairs, labelled A, B, and C, with coinciding ISFs (and then $\tau_{\alpha}$) and NGP; see Figure \ref{ISF+NGP} and References \cite{Tripodo_SM2019,Tripodo_epje_2019} for details. {Figure \ref{ISF+NGP} (left) shows that ISF has a characteristic plateau region at intermediate times. Increasing the relaxation time, the plateau widens, signalling that particles are  trapped for longer times by the surrounding particles (cage effect). The subsequent escape from the cage leads to the structural relaxation and  the ISF drop at $t \gtrsim \tau_{\alpha}$. The escape from the cage is strongly non-Gaussian, i.e., heterogeneous, as evidenced by the NGP growth around $\tau_{\alpha}$, revealing DH, i.e., the presence of particles with different mobility. }

		We now show how MI can be employed  to characterise DH. Additional details are given in References \cite{Tripodo_SM2019,Tripodo_epje_2019}. We consider the MI between the displacements of two generic particles $i$ and $j$
		\begin{equation}
			I_{ij}(t) = I(\delta \vec{r}_i(t),\delta \vec{r}_j(t)), \hspace{1cm} i \neq j
			\label{mi_disp}
		\end{equation}
		where $\delta \vec{r}_m(t)$ is the displacement of the $m$th particle within a time interval $t$, and MI from Equation (\ref{mut_inf}) is estimated by the Kraskov--St\"{o}gbauer--Grassberger (KSG) estimator \cite{Kraskov:2004qq}. The particles are said to be correlated at time $t$ if $I_{ij}(t) > I_0$ with $I_0 = 0.2$  \cite{Dunleavy:2015fq}.
		
\begin{figure}[h]
			\centering
			\includegraphics[width=0.8\textwidth]{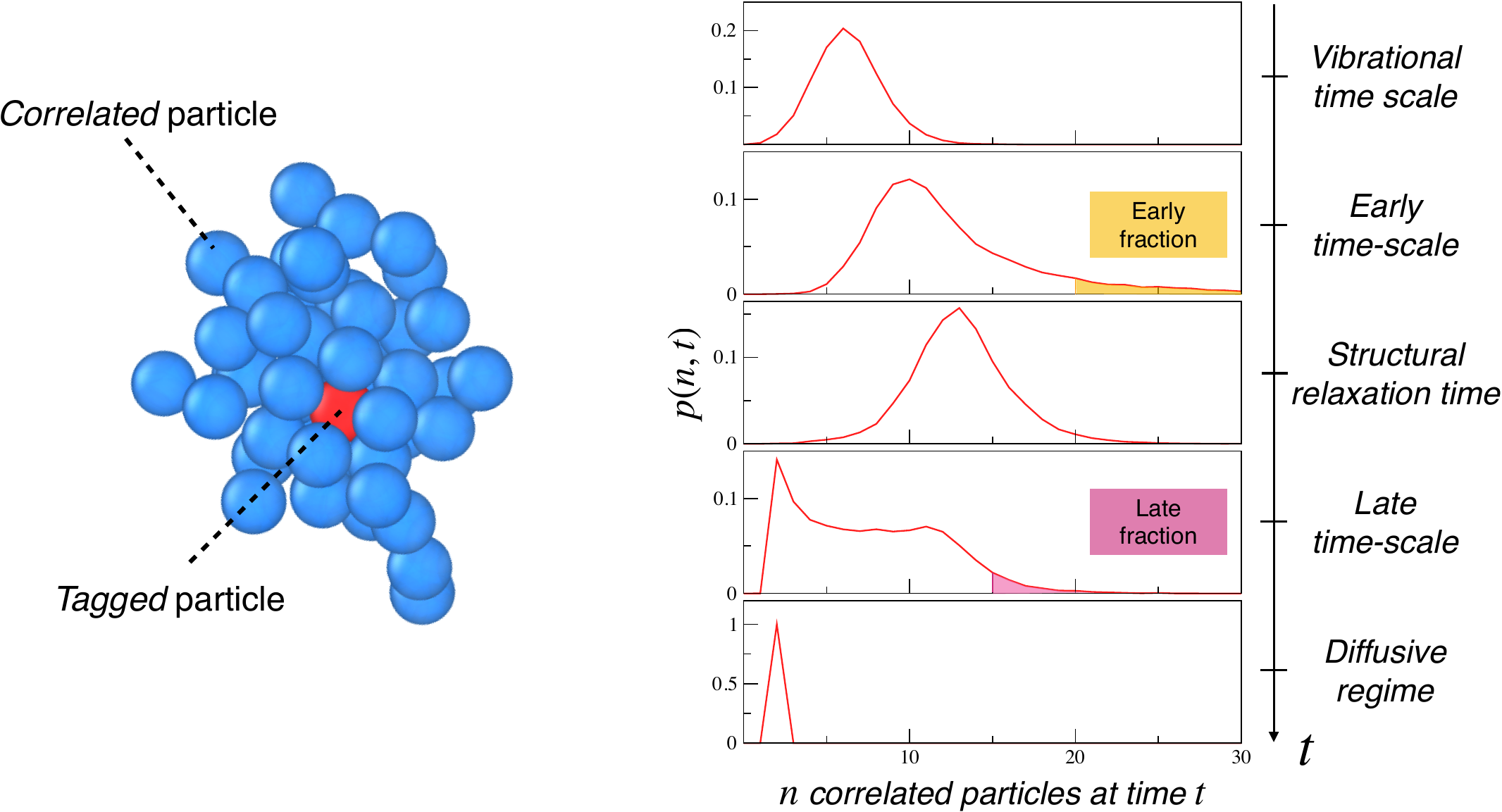}
			\caption{{
					(\textbf{Left}) Snapshot of the (blue-colored)  particles being MI-correlated to a (red-colored) tagged one. (\textbf{Right}) Time evolution of the distribution of the (red) particles correlated to  $n$ other (blue) particles at time $t$, $p(n,t)$, for a state of the C set (same shape for both states). Moving from short to long times, i.e., from the top to the bottom of the stacked plots, the second and the fourth plot correspond to $t = \tau_{early}$ and $t = \tau_{late}$, respectively, when the standard deviation $\sigma(t)$ of $p(n,t)$  is at a local maximum (see Figure \ref{AVG+STD}, right). The  fractions of particles with higher correlation at these two times are highlighted. They are referred to as \textit{early} and \textit{late} fractions \cite{Dunleavy:2015fq}. Their exact definition is given in main text. The data were taken from {Reference \cite{Tripodo_SM2019}.}}
			}
			\label{N_distrib}
		\end{figure}

		We define $n_i(t)$ as the number of particles that are MI-correlated to the generic $i$th at time $t$. The MI-correlated particles exhibit a tendency to group around the generic one; see Figure \ref{N_distrib}, left. We denote with $p(n,t)$ the distribution of particles being MI-correlated to other $n$ particles at time $t$. The distribution is evaluated over four {\it iso-configurational ensembles} (ICEs) of the liquid. Each ICE is represented by the particle displacements generated starting from the {\it same} initial overall particle configuration, assigning about $10^3$ different initial velocities to each particle according to the pertinent Maxwell distribution and evaluating the subsequent $10^3$ distinct time evolution of the liquid. It has been shown elsewhere that the average over the four ICEs are, within the errors, equivalent to the customary ensemble average, i.e., it adequately samples the phase space of the system \cite{Tripodo_SM2019}.
		
		Representative plots of $p(n,t)$ at different times are given in Figure \ref{N_distrib}, right. 
		At vibrational times, $t\sim 1$, the distribution peaks around the average number of surrounding particles hit by the tagged particle during the oscillatory motion within the cage where it is trapped. At longer times, still shorter than the structural relaxation time, $t \lesssim \tau_{\alpha}$, the tagged particle establishes further correlations through collision and  $p(n,t)$ spreads at higher $n$ values. At $t \simeq  \tau_{\alpha}$, when (on average) the central particle escapes from the cage, $p(n,t)$ narrows. Later, at $t \gtrsim \tau_{\alpha}$, the distribution widens again. Finally, at $t \gg \tau_{\alpha}$ in the diffusive regime, most correlation is lost and $p(n,t)$ approaches a time-independent narrow shape, peaking at the number of particles with permanent MI-correlation with the tagged particle, i.e., all of the particles but one belonging to the same molecule, $n=2$.  
		
		\begin{figure}[h]
			\centering
			\includegraphics[width=0.8\textwidth]{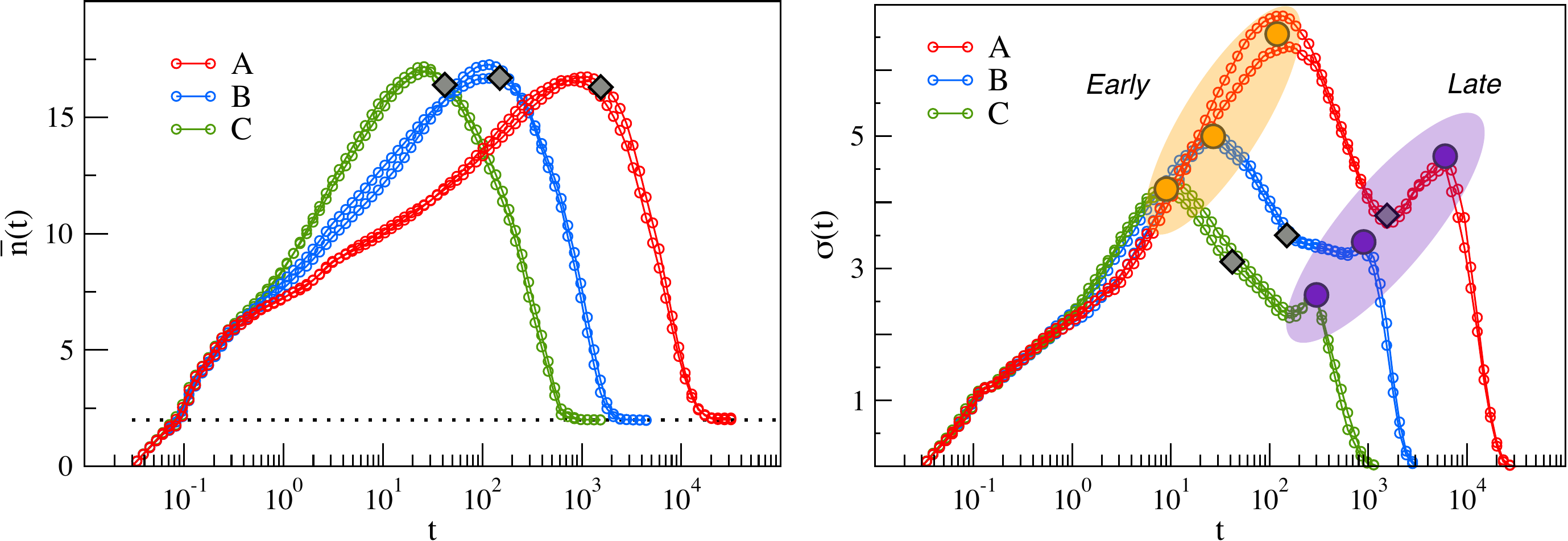}
			\caption{{Time}-dependence of $\bar{n}(t)$, the average number of particles that  are MI-correlated with a generic one (\textbf{left}), and $\sigma(t)$, the standard deviation (\textbf{right}) for all of the investigated states. Diamonds mark the position of $\tau_\alpha$. The two peaks of the standard deviation, also reported in atomic liquids \cite{Dunleavy:2015fq},  identify two time scales that are referred to as $\tau_{early}$ (orange dot) and $\tau_{late}$ (violet dot). The data were taken from Reference {\cite{Tripodo_epje_2019}}
			}
			\label{AVG+STD}
		\end{figure}

		Insight is provided by $\overline{n}(t)$ and $\sigma(t)$, the average and the standard deviation, respectively, of the number of MI-correlated particles  to a tagged one at time $t$. They are evaluated by $p(n,t)$. The results are shown in Figure \ref{AVG+STD}. Figure \ref{AVG+STD} (left) shows that the average number of MI-correlated particles $\overline{n}(t)$ reaches a peak at $t \simeq  \tau_{\alpha}$ and drops to $2$ at long times. Notably, {contrary to what happens in atomic liquids \cite{Dunleavy:2015fq}},  the height of the peak does {\it not} increase with $\tau_{\alpha}$ and the shapes of both $\overline{n}(t)$ and $\sigma(t)$ of states with equal $\tau_{\alpha}$ nearly coincide \cite{Tripodo_SM2019}.
		
		Intuitively, the standard deviation  $\sigma(t)$ offers an alternative DH metrics. Figure \ref{AVG+STD} (right) shows that it exhibits a peculiar bimodal structure similar to the one observed for the polydisperse mixture of hard spheres \cite{Dunleavy:2015fq}. Following Reference \cite{Dunleavy:2015fq}, we define two characteristic times, $\tau_{early}$ and $\tau_{late}$ ($\tau_{early} < \tau_{late}$),  corresponding to the position of the two maxima of $\sigma(t)$. In Figure \ref{N_distrib} right, the second and the fourth plots from the top are $p(n,\tau_{early})$ and $p(n,\tau_{late})$, respectively. We single out the fractions of particles  that are more correlated at  these two times; see the highlighted regions in Figure \ref{N_distrib} right. These fractions, henceforth referred to as \textit{early} or \textit{late}, respectively \cite{Dunleavy:2015fq},  correspond to the condition $n \ge \overline{n}(t) + 2 \, \sigma(t)$, $t = \tau_{early}, \tau_{late}$.   
		{Notably, the \textit{early} and \textit{late} fractions of particles show no intersection among them~\mbox{\cite{Tripodo_epje_2019,Dunleavy:2015fq}}}. The analysis of the mobility reveals (i) that the \textit{early} fraction has higher mobility and faster relaxation than the \textit{late} fraction~\cite{Tripodo_epje_2019} and (ii) that the time positions of the two peaks of the standard deviation are simply the relaxation time of the two fractions~\cite{Tripodo_epje_2019}. In addition to this, it has been verified that the noted scaling between fast vibrational dynamic and relaxation  observed in bulk systems (see for example Reference \cite{review_IJMS,OurNatPhys}) also holds for the \textit{early} and \textit{late} fractions~\cite{Tripodo_epje_2019}. Early and late fractions show a correlation with local \mbox{structure~\cite{Tripodo_epje_2019,Dunleavy:2015fq}}. In particular, the \textit{early} population has a tendency to be located in regions of the sample with lower local density and to be arranged in unstable topological structures. On the other hand, the \textit{late} one is correlated with the denser region of the sample and stable \mbox{topological structures}.

		\subsection{Clustering of \textit{Early} and  \textit{Late}  Fractions}
		\label{clustering}
		
		We now focus on the spatial distribution of the particles of the  \textit{early}  and \textit{late} populations and inspect the possible tendency to group in clusters. 
		
		The cluster analysis comprises three steps: (i) first, given a single microscopic configuration of a given state of the liquid, the particles that belong to the \textit{early} or \textit{late}  fractions are identified according to the ICE procedure outlined in Section \ref{previousDH}; (ii) then, within each fraction, clusters of particles are searched according to their positions in the given configuration; and (iii) finally, suitable averages over four distinct initial configurations are performed to provide statistical significance.
		
		{To detail the second part of the cluster analysis, we start by defining a cluster as a set of $M$ particles of the same population ($M\ge2$) with the distance from at least one other member of the cluster being less than $\ell$. We choose $\ell$ as the first minimum of the radial distribution function, which, depending on the sample and the state, lies in the range $1.44\leq \ell \leq1.46$. Within this range, the exact value of $\ell$ does not affect the results.} The clusters are identified by a bottom-up algorithm performing the following steps  \cite{BottomUpAlgorithm}:
		
		\begin{enumerate}
			\item A particle of a given population is chosen and included as first member of a\linebreak \mbox{possible cluster;}
			\item Particles of the same population within distance $\ell$ from the first member are searched and, in the positive case, added to the cluster;
			\item Step 2 is repeated for all {\it new} members of the cluster;
			\item If no new members are found and $M\ge2$, the cluster is completed and its size \mbox{$M$ is defined; }
			\item All of the particles involved in the already identified clusters are removed, and the procedure is restarted from step 1 by considering one left particle. 
			
		\end{enumerate}
		
\begin{figure}[h]
			\centering
			\includegraphics[width=0.5\textwidth]{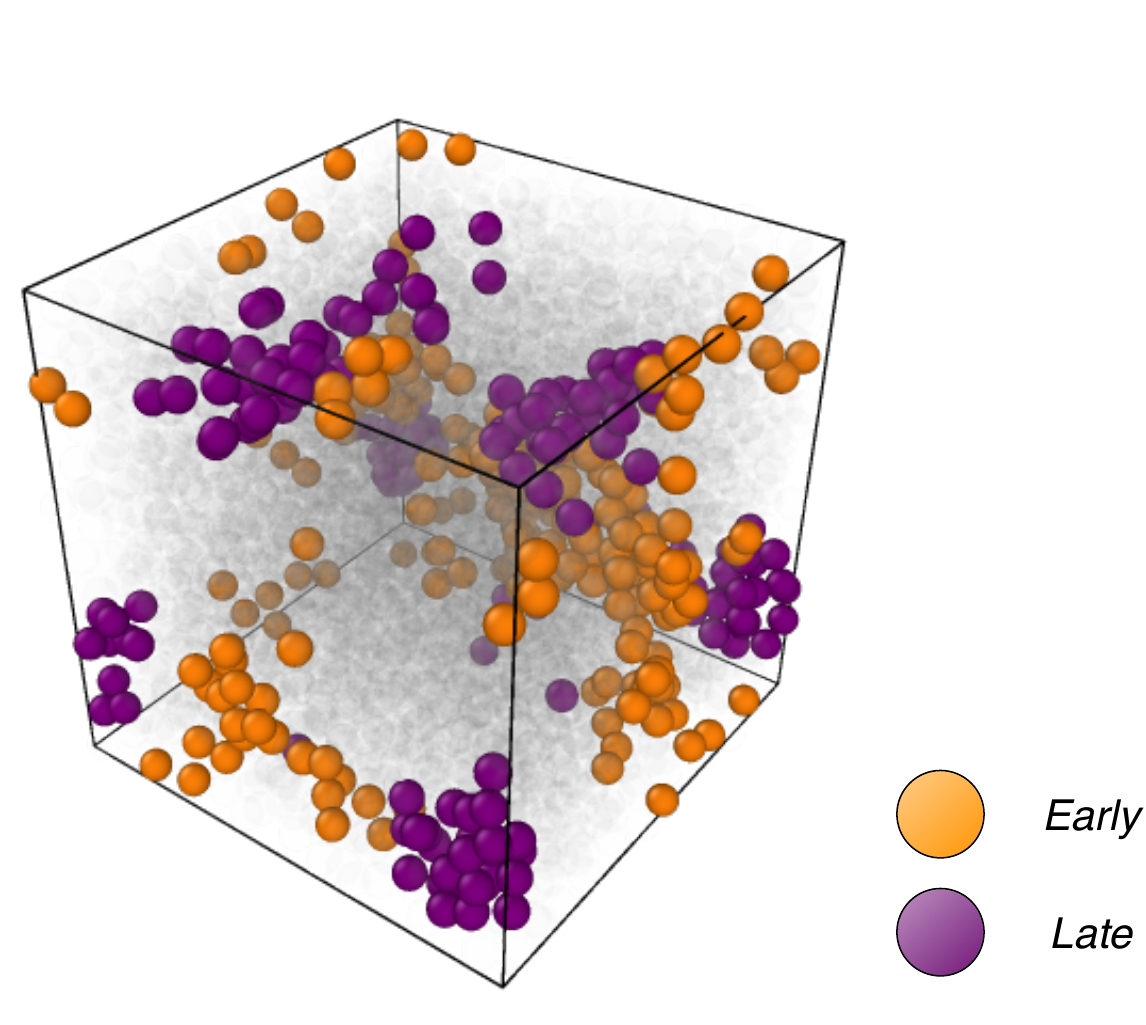}
			\caption{{
					View of one microscopic configuration of the state with $\rho=1.02$ and $T=0.42$ belonging to the C set. The particles that belong at the \textit{early} and \textit{late}  fractions are highlighted. Note the tendency of the particles belonging  to the two fractions to group into clusters coexisting with {\it isolated} particles .
			}}
			\label{cluster_fig}
		\end{figure}

		Figure \ref{cluster_fig} provides an example of the outcome of the first and second steps of the cluster analysis for a single initial configuration. 
		
\begin{figure}[h]
			\centering
			\includegraphics[width=0.8\textwidth]{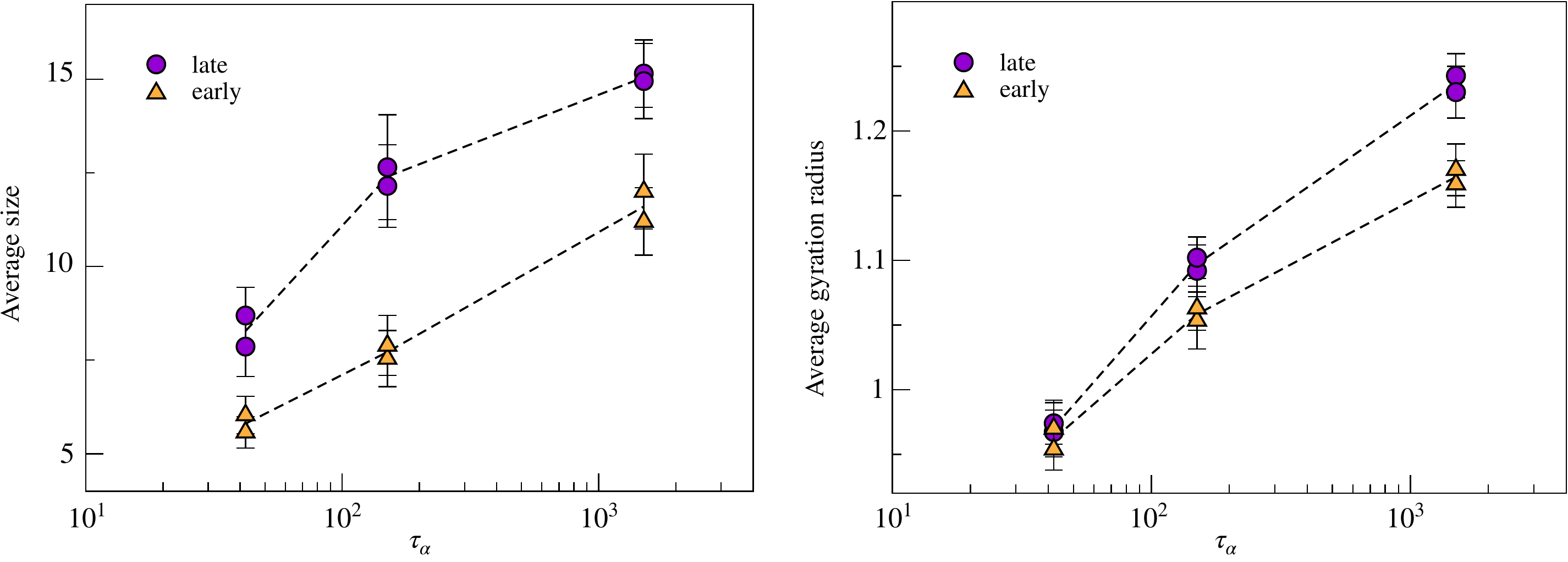}
			\caption{{Cluster analysis of the early and late populations of the three pairs of states with equal structural relaxation times $\tau_\alpha$. (\textbf{Left}) Average number of particles forming a cluster. (\textbf{Right}) Average cluster gyration radius. The clusters of both populations increase in both size and gyration radius with the relaxation time. Notably, the clusters of states with {\it equal} relaxation times have {\it equal} average size and gyration radius within the errors.}}
			\label{M+RG}
		\end{figure}

		After completion of the cluster analysis by performing the third step, i.e., the average over the four ICEs, we start to characterise the clusters of particles belonging to the  \textit{early}  and \textit{late} populations. To this aim, we first consider the average cluster size and the corresponding gyration radius $R_g$. The latter is defined for a cluster with size $M$ as 
		\begin{equation}
			R_g^{cl}=\left[\frac{1}{M} \sum_{i\in cl} ({\bold r}_i-{\bold r}_{cm})^2 \right]^{\frac{1}{2}}
		\end{equation}
		where ${\bold r}_{cm}$ is the cluster center of mass.  The results are plotted in  Figure \ref{M+RG}.  They show a moderate increase of both the average size and radius with increasing relaxation time, which is fully consistent with the increasing DH as signalled by NGP; see Figure~\ref{ISF+NGP}, right.  Notably, the clusters of states with {\it equal} relaxation times have {\it equal} average size and gyration radius within the errors.

\begin{figure}[h]
			\centering
			\includegraphics[width=0.8\textwidth]{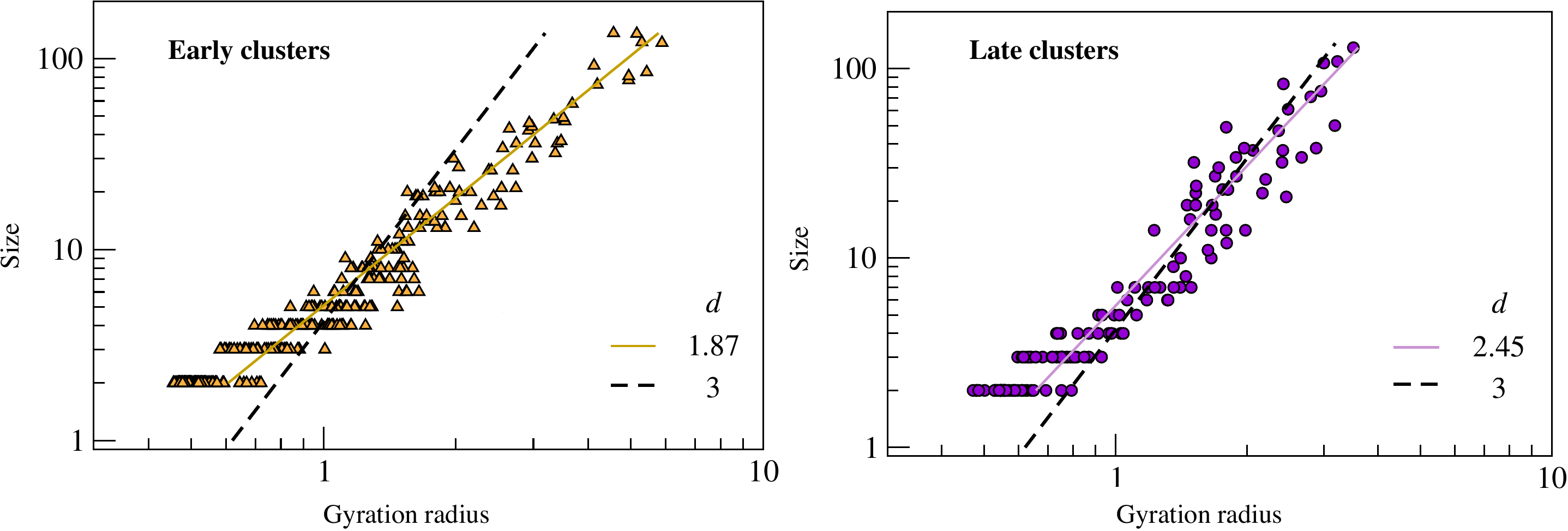}
			\caption{{Correlation plot between the size $M$ and the gyration radius $R_g$  of all clusters found in the \textit{early} (\textbf{left}) and \textit{late} (\textbf{right}) fractions of the six states of the liquid under study. The black dashed lines are the best-fit curves according to Equation (\ref{dimensionality}) with $d=3$ by adjusting the parameter $c$ only. The continuous lines are the best-fit curves according to Equation (\ref{dimensionality}) by adjusting also the fractal dimension $d$. We find $d=1.87(2)$ and $d=2.45(3)$ for the \textit{early} and \textit{late} clusters, respectively.}}
			\label{early_late_scatter}
		\end{figure}

		Further insight is provided by the analysis of the fractal dimension  $d$ of the clusters that is drawn by the relation between the size $M$ and the gyration radius
		\begin{equation}
			M=c \, R_g^{d}
			\label{dimensionality}
		\end{equation}
		where $c$ is a suitable constant. Figure \ref{early_late_scatter} shows the good correlation between the size and the gyration radius of all of the  clusters in all states considered in this work. A best-fit of the data by using Equation (\ref{dimensionality}) yields  $d=1.87(2)$ and $d=2.45(3)$ for the \textit{early} and \textit{late} clusters, respectively. The results suggest that the clusters of the \textit{early} fractions are more filamentous, reminding us of the low-dimensional structures with correlated motion observed in glassforming liquids \cite{Donati_PRL1998,Donati_PRL1999,Starr_JCP2013}.  Conversely, \textit{late} particles appear to be arranged in compact, sphere-like clusters with $d=2.45(3)$. These conclusions are strengthened by limiting the fit procedure to the largest clusters with $M\ge3$ and $M\ge5$, for which
		One finds $d=1.96(2)$ and $d=2.00(1)$ for the early fraction, respectively, but for which, one finds $d=2.63(2)$ and $d=2.98(2)$ for the late
		fraction, respectively.

		\section{Mutual Information and Johari--Goldstein  $ \beta$-Relaxation in a Model Polymer Melt}
		
		When approaching GT, molecular rearrangements occur via both primary modes, referred to as structural or $\alpha$ relaxation, and via the faster secondary ($\beta$) processes, as evidenced by the mechanical, electrical, and thermal properties of the materials  \cite{WilliamsMcCrum, AngelNgai00,NgaiBook}.
		Although it has been the topic of a large number of phenomenological and theoretical studies as well as of experiments and simulations
		\cite{JOHARI70, Ngai98, NgaiPaluchClassificationSecondaryJCP04, NgaiBook, Capaccioli12, Cicerone2014,yu2017}, there is still no definitive microscopic description available for the $\beta$ relaxation.
		There is a special class of secondary relaxations involving the translation or reorientation of the molecular unit as a whole, as opposed to the trivial ones that are usually related to intramolecular degrees of freedom, such as the motion of pendant groups in polymers. 
		This special class of $\beta$ processes, called Johari--Goldstein (JG), to honor the researchers that first noticed it \cite{JOHARI70}, is universal in glass \mbox{formers \cite{Ngai98, NgaiPaluchClassificationSecondaryJCP04, NgaiBook}} and argued to be the precursor of the structural relaxation \cite{Ngai98, Capaccioli12,NgaiPaluchClassificationSecondaryJCP04, NgaiBook}. 
		Recently, the investigation of JG relaxation using MD simulations has been facilitated by the development of coarse-grained models of linear polymers overcoming the need for complex chain \mbox{architectures \cite{BedrovPRE2005,BedrovJNCS2011,FragiadakisRolandMM17}}. It is also worth mentioning the studies of the JG relaxation in asymmetric diatomic \mbox{molecules \cite{FragiadakisRolandPRE12,FragiadakisRolandMM17}.}
		
\begin{figure}[t]
			\centering
			\includegraphics[width=0.7\textwidth]{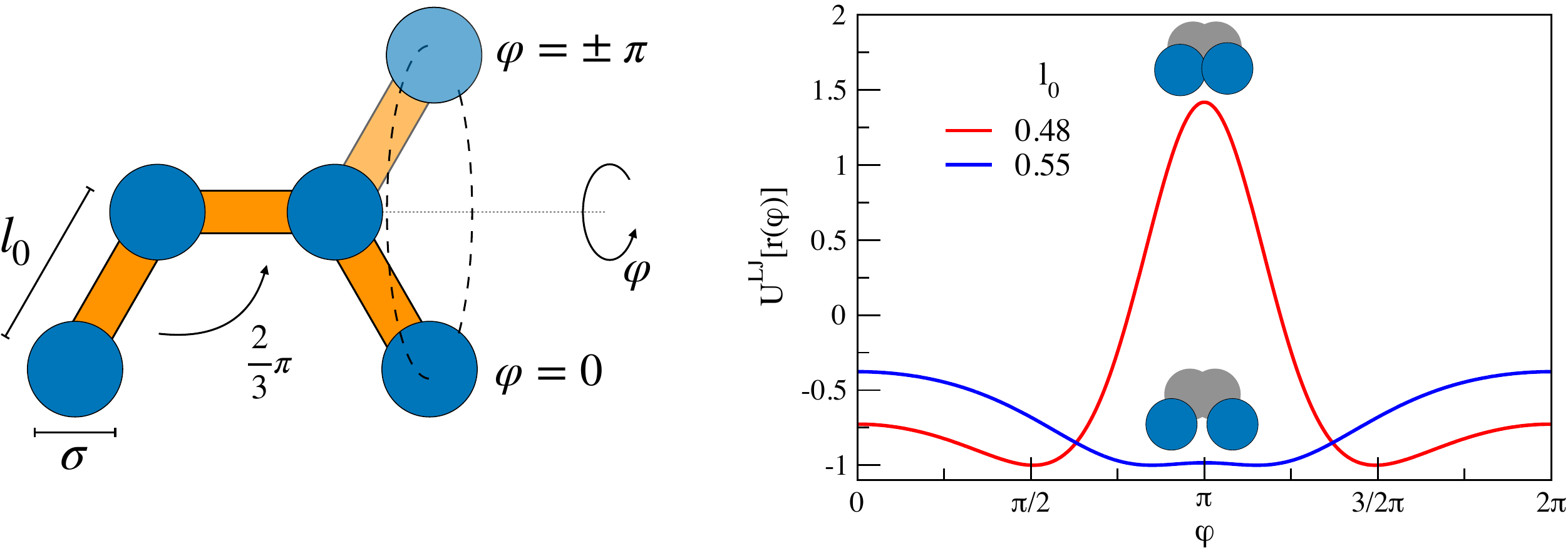}
			\caption{(\textbf{Left}) {Schematic} representation of a fragment of the polymeric chain model adopted in the MD simulations. For clarity reasons, the bond lengths are exaggerated. The dihedral angle $\varphi$ is the angle by the two planes defined by the first and last three monomers of the fragment. (\textbf{Right}) Effective torsional potential resulting from the Lennard--Jones interaction between the first and the last monomers of the fragment. An abrupt change in the potential occurs by shortening the bond length $l_0$ from $0.55 \sigma$ to $0.48 \sigma$ with the appearance of a considerable energy barrier.}
			\label{bond}
		\end{figure}

		We performed extensive MD simulations of a model linear polymer with nearly fixed bond length $l_0$ and bond angles constrained to $2\pi/3$; see Figure \ref{bond} left. We considered $N_c = 512$ linear chains made of $M = 25$ monomers each, resulting in a total number of monomers $N = 12800$. 
		Further details are given elsewhere \cite{Tripodo_polymers2020,PuosiTripodo_Macromolecules2021}.  An interesting aspect of the model is that, even if the force field does not include a torsional interaction, an effective torsional barrier may hinder the dihedral rotations depending on the bond length $l_0$ due to the LJ repulsion between two non-adjacent monomers. We studied the two cases $l_0=0.48 \,\sigma$ and $l_0=0.55 \,\sigma$, which present  considerable or missing torsional barrier, respectively; see Figure \ref{bond} right. Henceforth, all lengths are expressed in units of $\sigma$.
		
		\subsection{Mutual Information and Bond Reorientation}
		\label{subsection_mi_bond}
		
		Since in linear chains the JG process involves local motion of the polymer chain backbone \cite{NgaiPaluchClassificationSecondaryJCP04, NgaiBook}, we focus on the most elementary relaxation process, that is, the reorientation of the bond linking two adjacent monomers of the same chain. 
		To this aim, we consider the unit vector along the $m$th bond of the $n$th chain at time $t$
		\begin{equation}\label{eq:bmn}
			{\bf b}_{m,n}(t) =\frac{1}{l_0}({\bf r}_{m,n}(t)-{\bf r}_{m+1,n}(t)),    
		\end{equation}
		where ${\bf r}_{m,n}(t)$  denotes the position of the $m$th monomer in the $n$th chain at time $t$. Hence, we define the bond correlation function (BCF) $C(t)$ \cite{CapacciEtAl04}
		\begin{equation}
			C(t)= \frac{1}{N_c} \frac{1}{M-1} \sum_{n=1}^{N_c} \sum_{m=1}^{M-1}  \langle {\bf b}_{m,n}(t)\cdot{\bf b}_{m,n}(0) \rangle.
			\label{Cbond}
		\end{equation}

		The brackets $\langle \cdots \rangle$ denote a suitable ensemble average. { It is worth noting that BCF is a kind of  Pearson correlation and then senses {\it linear} dependencies only. }
		Figure \ref{bcf} plots BCF of the two polymer melts with chains having different bond length. In agreement with previous studies \cite{BedrovJNCS2011,BedrovPRE2005},  we found that the chains having shorter bond length exhibits a characteristic two-step decay after a first fast decay at $t \sim 0.1$, which is a signature of the presence of two distinct relaxation processes. We ascribed the faster one to the JG $\beta$ relaxation and the slower one to the structural $\alpha$ relaxation \cite{Tripodo_polymers2020,PuosiTripodo_Macromolecules2021}.
		
		\subsubsection{MI Correlation in Time Domain}
		
		We now show that replacing BCF from Equation (\ref{Cbond}) with the corresponding MI quantity provides sharper information on the bond reorientation and, in turn, on JG relaxation. This suggests that the JG process has significant {\it nonlinear} contributions, as anticipated in a dense glassformers \cite{Ngai00,BordatNgai04}.  The MI correlation of the bond reorientation is expressed by the quantity $I({\bf b}(t+t_0),{\bf b}(t_0))$, i.e., Equation (\ref{mut_inf}) with $X={\bf b}(t)$, the bond orientation at time $t$, and with $Y={\bf b}(0)$, the bond orientation at the reference time $t_0$. For simplicity we set $t_0=0$.
		\begin{figure}[h!]
			\centering
			\includegraphics[width=0.8\textwidth]{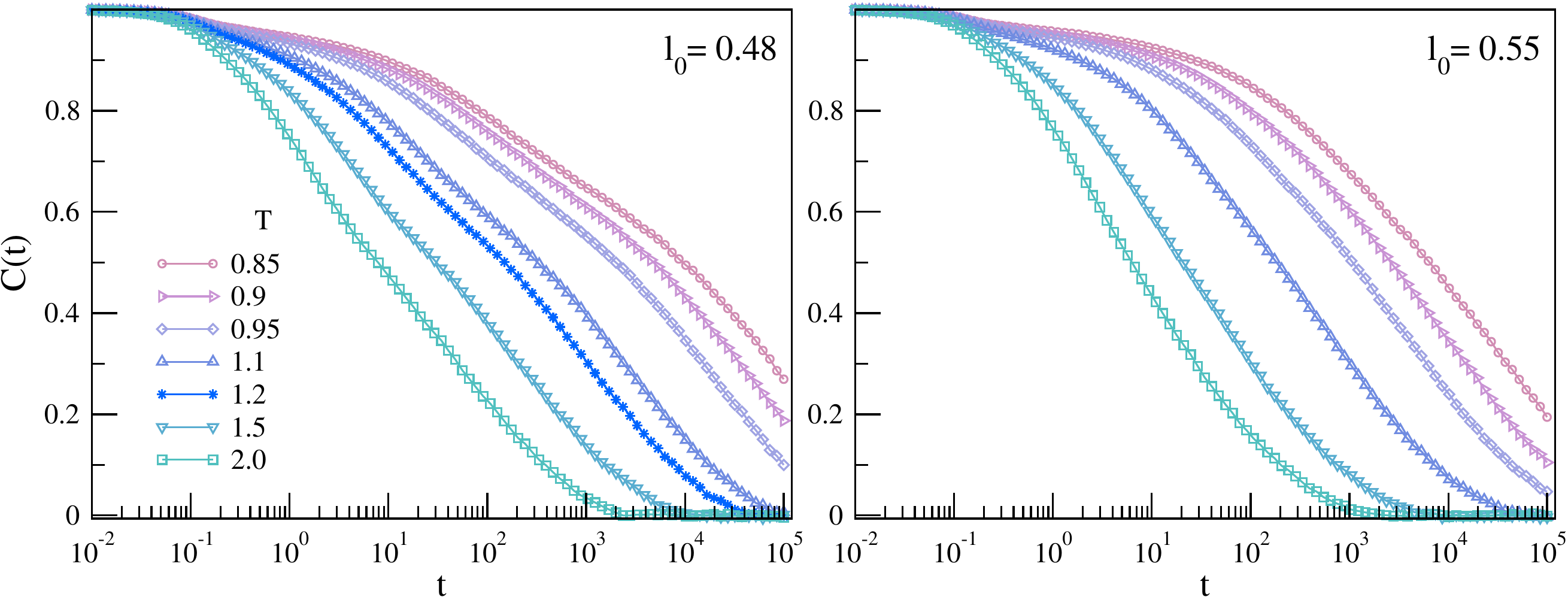}
			\caption{{Temperature} dependence of the BCF of the chains with different bond lengths. If $l_0=0.48$, a two-step decay--evidencing two distinct relaxations--is observed. We ascribed the faster one to the JG $\beta$ relaxation and the slower one to the structural $\alpha$ relaxation \cite{Tripodo_polymers2020,PuosiTripodo_Macromolecules2021}. The data were taken from Reference {\cite{Tripodo_polymers2020}}
				.}
			\label{bcf}
		\end{figure}

\begin{figure}[h]
			\centering
			\includegraphics[width=0.8\textwidth]{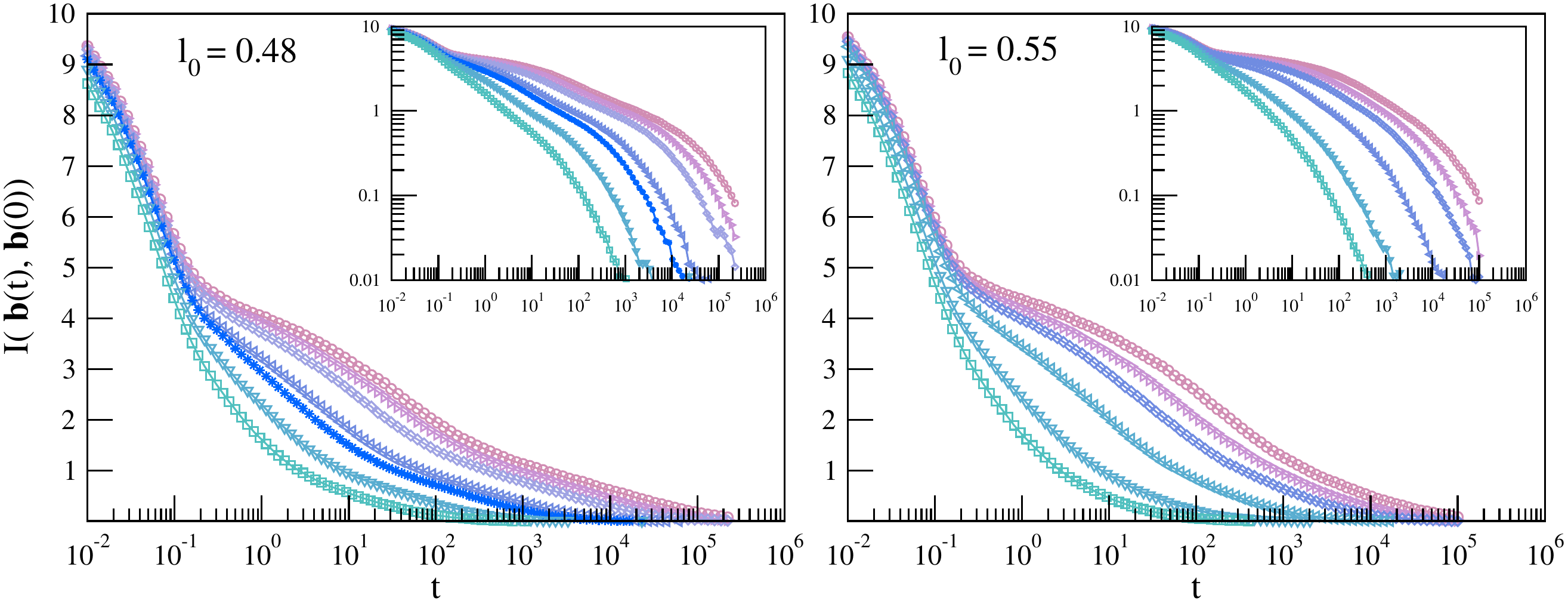}
			\caption{{Temperature} dependence of the MI correlation loss between the bond orientations at $t_0$ and $t_0+t$  of the chains with different bond lengths. The color codes are as in Figure \ref{bcf}.}
			\label{mi_rotation}
		\end{figure}

\begin{figure}[h]
			\centering
			\includegraphics[width=0.5 \textwidth]{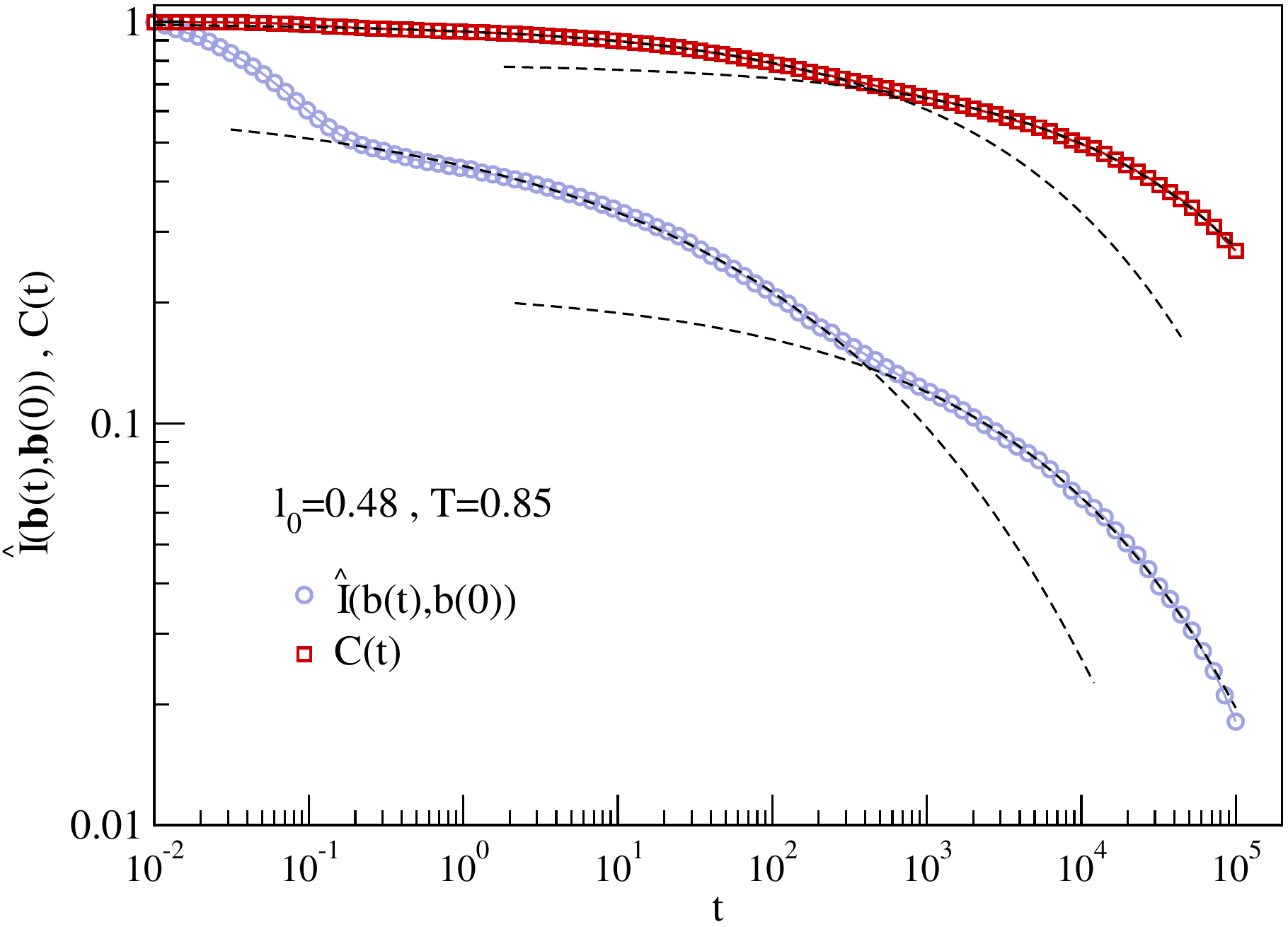} %
			\caption{{Comparison} of the BCF $C(t)$ with the corresponding MI normalised quantity $\hat{I}({\bf b}(t),{\bf b}(0))$ for the state with slowest relaxation. The superimposed dashed lines are guides for the eyes. The better resolution of MI to the JG relaxation is quite apparent.}
			\label{comparison_mi-bcf}
		\end{figure}

		Figure \ref{mi_rotation} plots $I({\bf b}(t),{\bf b}(0))$ for the polymer melts with (left panel) and without (right panel) JG relaxation. In both cases, it is seen that about $50$--$60 \%$ of MI is lost quite fast within $t \sim 0.1$, corresponding to a few picoseconds. This correlation loss is significantly higher than in the case of  BCF; see Figure \ref{bcf}. Nonetheless, MI gives a hint of  a clearer multiple decay pattern at intermediate and long times  in the presence of JG relaxation. 
		The increased resolution of MI to JG relaxation is substantiated for the state with slowest relaxation by Figure \ref{comparison_mi-bcf} comparing BCF and the normalised MI correlation function:  
		\begin{equation}
			\hat{I}({\bf b}(t),{\bf b}(0))= I({\bf b}(t),{\bf b}(0))/I({\bf b}(0),{\bf b}(0))
			\label{MInorm}
		\end{equation}
		\vspace{-12pt}
		
		\subsubsection{MI Correlation in the Frequency Domain}
		
		Having shown the increased resolution of MI  to JG relaxation with respect to BCF in the {\it time} domain, it is legitimate to wonder how this is reflected in the {\it frequency} domain. Here, the quantity of interest is the spectral response function, or susceptibility, to be defined as
		\begin{equation}
			\chi_m(\omega)=\chi_m'(\omega)+i\chi_m''(\omega)=1+i\omega\int_{0}^{\infty}dt\;e^{i\omega t}\Phi_m(t), \hspace{1cm} m = BCF, MI
			\label{chi}
		\end{equation}
		Our plan is to compare $\chi_{BCF}(\omega)$ and $\chi_{MI}(\omega)$, i.e., Equation (\ref{chi}) with $\Phi_{BCF}(t)= C(t)$ (Equation (\ref{Cbond}))  or with $\Phi_{MI}(t)= \hat{I}({\bf b}(t),{\bf b}(0))$ (Equation (\ref{MInorm})), respectively. 
		
\begin{figure}[h]
			\centering
			\includegraphics[width=0.7 \textwidth]{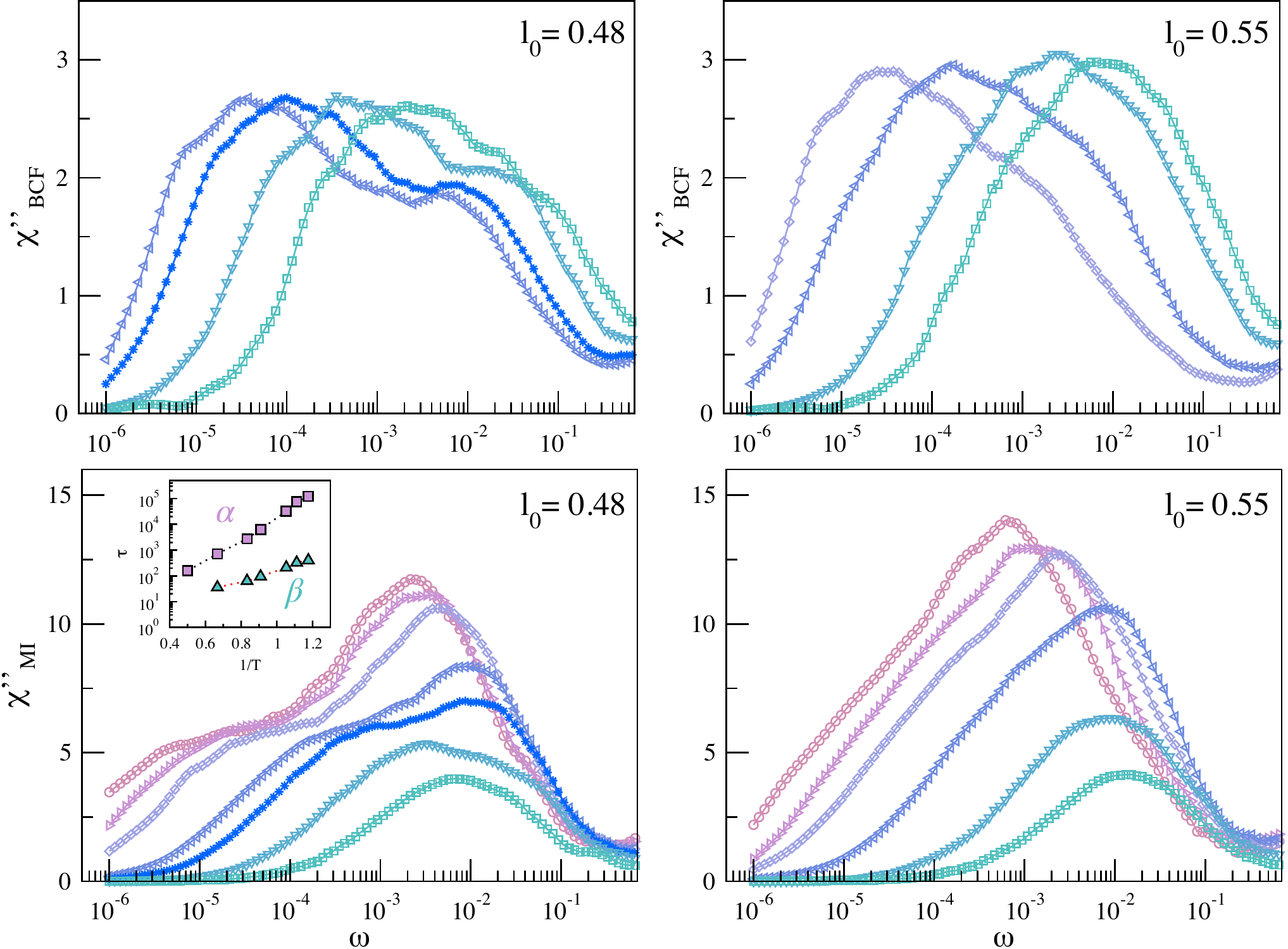}
			\caption{{{Temperature} dependence of the imaginary part of the susceptibility $\chi''_{BCF}(\omega)$  (\textbf{top}) and  $\chi_{MI}(\omega)$ (\textbf{bottom}) for the systems with ($l_0=0.48)$ and without  ($l_0=0.55$) JG relaxation. Inset: Arrhenius plot of the quantity $\tau_i = 1/\omega_{i, max}$ with $\omega_{i, max}$, $i= \alpha, \beta$, and the frequency of the local maxima of  $\chi_{MI}(\omega)$ ($\omega_{\alpha, max} < \omega_{\beta, max}$). The color codes are as in Figure \ref{bcf}.}}
			\label{comparison_chi_sec}
		\end{figure}

		Figure \ref{comparison_chi_sec} compares the imaginary part of the susceptibility,  $\chi''_{BCF}(\omega)$ and $\chi''_{MI}(\omega)$ for the systems with and without JG relaxation. It is seen that the presence of JG relaxation is weakly detectable in $\chi''_{BCF}(\omega)$. Unfortunately, given the investigated time window, due to the slow decay rate of BCF, it is not possible to properly compute $\chi''_{BCF}(\omega)$ down to the lowest studied temperature.
		Differently, a well-defined double-peaked pattern of $\chi''_{MI}(\omega)$, ascribed to splitting of the two peaks associated to the $\alpha$ and $\beta$ processes, develops when decreasing the temperature. The inset shows the Arrhenius plot of the two time scales, $\tau_i = 1/\omega_{i, max}$, $i= \alpha, \beta$, with $\omega_{\alpha, max}$ and $\omega_{\beta, max}$ being the local maxima of $\chi''_{MI}(\omega)$,  ($\omega_{\alpha, max} < \omega_{\beta, max}$). The maxima have been searched by best-fitting $\chi''_{MI}(\omega)$ with a weighed sum of two Gaussians. On cooling, the increasing splitting of the time scales $\tau_\alpha$ and $\tau_\beta$ of the  primary and the secondary JG relaxation is apparent.
		\newpage
		\subsection{Dynamic Heterogeneity and Mutual Information between Displacement and Rotation of \mbox{the Bond}}
		
		{ 
			In the polymer model under study, Figure \ref{bond}, a shorter bond length yields an effective torsional barrier. We argued that the role of the barrier is twofold \cite{PuosiTripodo_Macromolecules2021}: (i) it is responsible for a partial averaging of the DH at the JG relaxation time scale, giving rise to a peculiar double-peaked time evolution of the NGP with two local maxima located  at the JG and primary relaxation time scales, respectively, and (ii)  it slows down the bond {\it reorientation} and, jointly, the monomer {\it translation}, thus suggesting a roto-translation coupling, which, indeed, has been recently reported in glassforming systems with strong JG relaxation \cite{CapacciJGRotoTranslSciRep19}.
			
			Here, we expose the roto-translation coupling by quantifying the correlation between the  bond {\it reorientation} and the monomer {\it translation}. We find that the correlation is higher at the time scales of the primary and JG secondary relaxation when DH is also at a \mbox{local maximum.}
		}
		
\begin{figure}[h]
			\centering
			\includegraphics[width=0.3\textwidth]{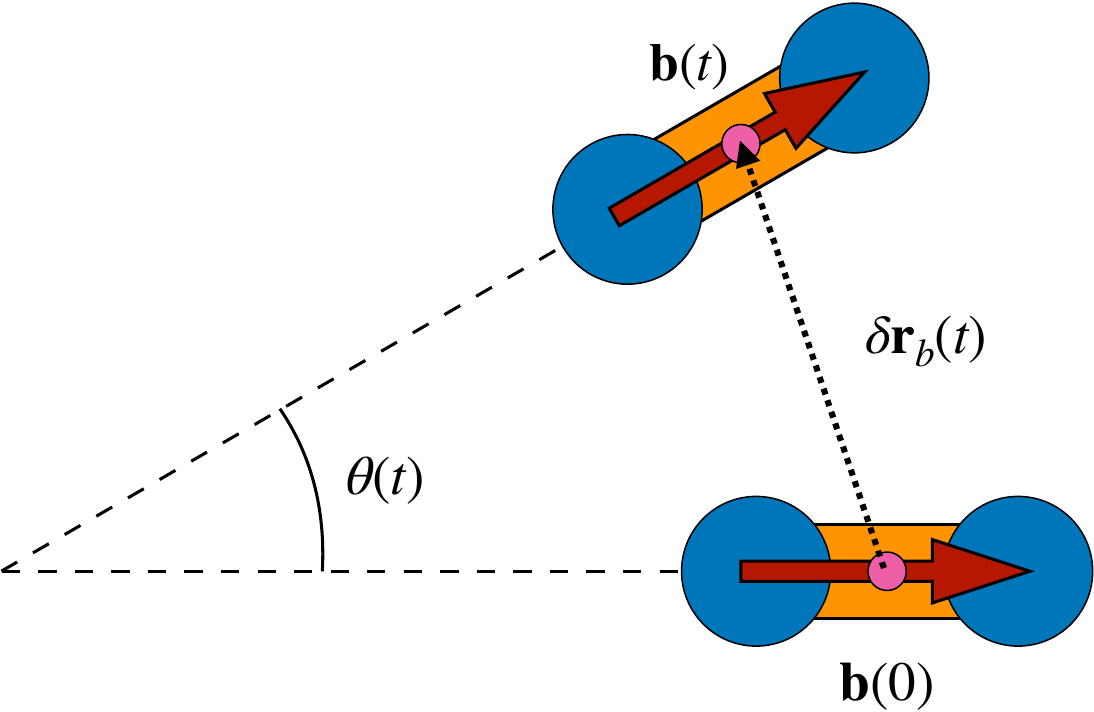}
			\caption{Schematic representation of the angle covered by a given bond in a time interval $t$, $\theta(t)$, and $\delta {\bf r}_{b}(t)$, the displacement of the bond position, i.e., the center of mass of the two monomers linked by the bond.}
			\label{cartoon_rot-disp}
		\end{figure}

		To provide evidence, we consider the bond reorientation in a time $t$, $\theta(t)$, and the corresponding squared modulus, $\delta r^2_{b}(t))$, of the bond displacement, $\delta {\bf r}_{b}(t)$, i.e., the displacement of the center of mass of the two monomers linked by the bond;  {see Figure \ref{cartoon_rot-disp}}. We evaluate the equal-time Pearson correlation $C(\cos\theta(t),\delta r^2_{b}(t))$ and the analogous MI correlation according to Equations (\ref{Pearson_def}) and (\ref{mut_inf}) with $X=\cos\theta(t)$ and $Y=\delta r^2_{b}(t)$.
		
\begin{figure}[h]
			\centering
			\includegraphics[width=0.7\textwidth]{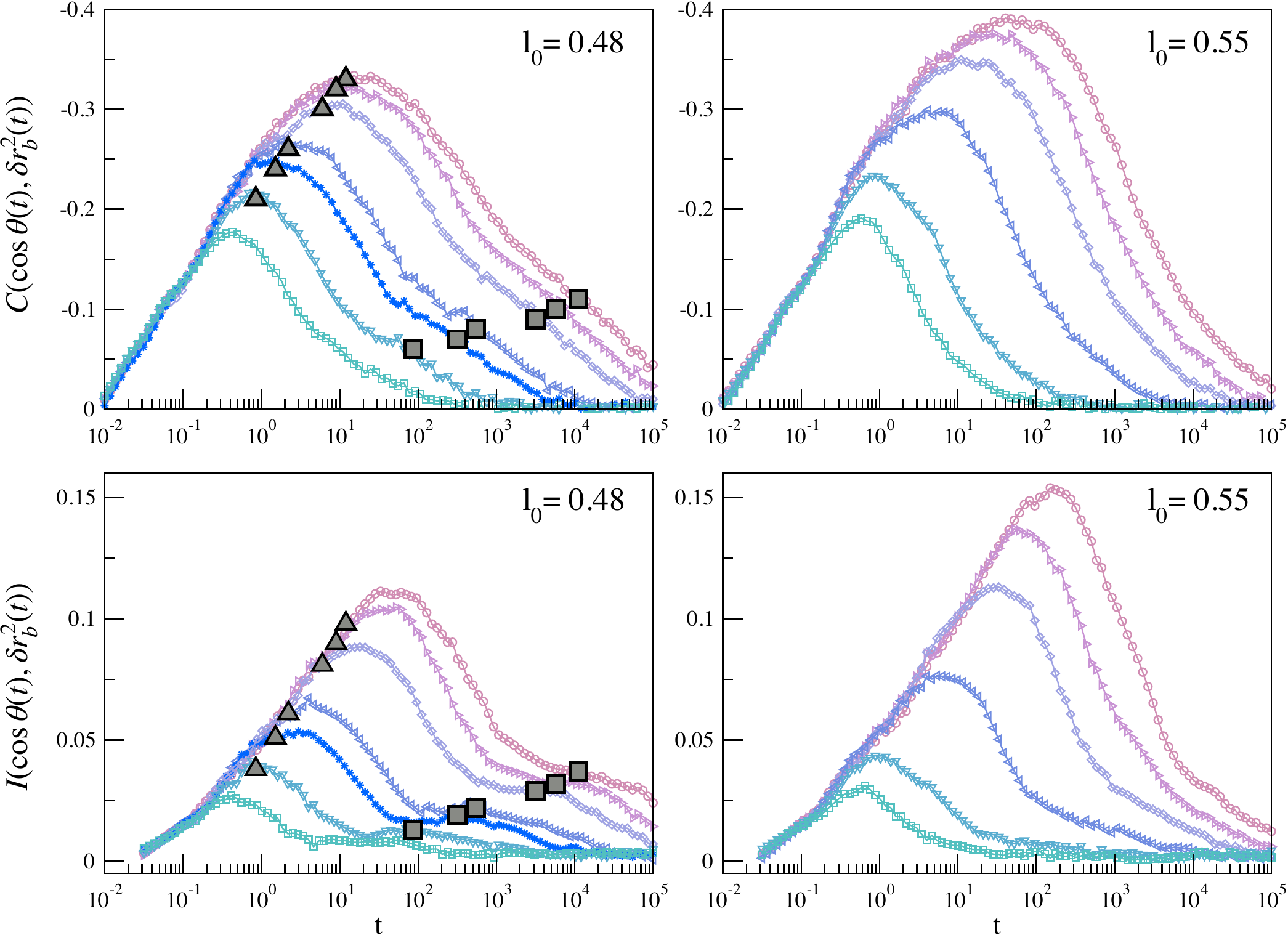}
			\caption{{{Time} dependence of the equal-time correlation between the reorientation and the square displacement of a given bond in a time interval $t$ according to Pearson (\textbf{top}) and MI (\textbf{bottom}) for the systems with ($l_0=0.48)$ and without  ($l_0=0.55$) JG relaxation. The dots indicate the position of the NGP local maxima at the JG (triangle) and structural (square) relaxation time scales {\cite{PuosiTripodo_Macromolecules2021}}
				}. {Color codes as in Figure \ref{bcf}.}}
			\label{correlation_rot-disp}
		\end{figure}

		Figure \ref{correlation_rot-disp} plots the above quantities. 
		The correlation grows with time up to the JG time scale and then decays at long times in the presence of structural relaxation.  In the presence of JG relaxation, two major facts are apparent. First, MI reveals a bimodal structure peaking at the time scales of JG and $\alpha$ processes. The structure is quite attenuated by Pearson correlation. Second, MI correlations reach a local maximum when the NGP reaches a maximum too. The finding points to the conclusion that the primary and the secondary relaxations have in common not only the DH presence, as already noted elsewhere \cite{PuosiTripodo_Macromolecules2021}, but also the significant correlation between the bond reorientation and displacement. This suggests that, in molecular liquids, the mechanistic explanation of both DH and relaxation involves the rotation/translation coupling.
		
		\section{Conclusions}
		
		We reported on extensive MD simulations concerning two coarse-grained models of dense fluids, a molecular liquid and a polymer melt. The focus was on the assessment of the additional insight offered by MI with respect to the usual Pearson correlation when dealing with two major features of transport and relaxation close to GT, namely DH and JG secondary relaxation. 
		
		In the molecular liquid, significant DH was evidenced. MI detected two distinct fractions of particles, so-called \textit{early} and \textit{late} fractions, which were spatially grouped in clusters with filamentous and compact structures, respectively, and exhibited different mobility and relaxation properties.
		
		In the polymer melt, the JG relaxation was tuned by the bond length and was better revealed by MI analysis. We found that both DH and MI between the reorientation and the displacement of the bonds reach their (local) maxima at the time scales of the primary and JG secondary relaxations. This suggests that, in (macro)molecular systems, the mechanistic explanation of both DH and relaxation must involve the rotation/translation coupling.
		
		\paragraph{Acknowledgements} 
		Generous grants of computing time from the Green Data Center of the University
		of Pisa and from Dell EMC$^{\circledR}$ Italia are also gratefully acknowledged.
		
		\bibliography{biblio}
		
		\appendix

	\end{document}